# Anatomy of a Spreadsheet Failure: Analysing the EuSpRIG Horror Story Corpus

Simon Thorne and Angela Collins
Cardiff Metropolitan University and Dow Scholfield Watts (DSW)
sthorne@cardiffmet.ac.uk, angela@dswbusinessplanning.com

**ABSTRACT**

Spreadsheet disasters have been documented for more than thirty years, but they have not gone away. This paper examines the EuSpRIG horror stories archive as evidence of how ordinary spreadsheets fail and the consequences that follow. After cleaning and deduplication, 121 incidents from 1995 to 2026 were classified into twelve failure categories and analysed by mechanism, context and consequence. Formula errors, data-entry slips and data-handling mistakes account for 71 cases and recur across the whole period, from Fidelity's missing minus sign in 1995 to a mistyped date that cost Norway's sovereign wealth fund about $92 million in 2024. A second pattern concerns information present in a workbook but not visible to the recipient: hidden rows, hidden sheets, pivot-table caches, embedded objects and retained source layers can cause harm even when no calculation is wrong. The clearest rise in reported cases is in this hidden or embedded data category, whose most serious example is the 2022 Ministry of Defence spreadsheet that exposed details of roughly 18,700 Afghan applicants through hidden rows. The paper argues that spreadsheet risk is no longer only a problem of wrong numbers. It also involves visibility, disclosure, governance and trust. Existing taxonomies remain useful for formula, entry and handling errors, but do not fully capture the public harm caused by spreadsheets used as uncontrolled operational infrastructure. The same mistakes have persisted for three decades; what has changed is the authority organisations give to the spreadsheets that contain them.

## 1.0 INTRODUCTION

Around 1800 BC, a scribe in southern Mesopotamia pressed a table of numbers into a clay tablet now known as Plimpton 322. It was not a spreadsheet in the modern sense, but it used a familiar tabular structure, rows and columns of related values, set out so that each can be read against the others. It is also, in at least a couple of places apparently incorrect, the tablet contains transcription slips: a digit miscopied and a value that does not reconcile. Nearly four thousand years before Excel, the oldest known numerical table already carried the oldest known table errors.

In January 1995 a tax accountant at Fidelity's Magellan fund sat down to estimate the year's dividend. The fund had realised a $1.3 billion net capital loss, and the figure went into a spreadsheet omitting its minus sign. The loss became a gain, the estimated distribution swung by some $2.6 billion, and the fund had to write to shareholders to take back a dividend it had already promised them [NS13] (Hinden, 1995). Twenty-nine years later, in January 2024, the largest pool of investable capital on earth, Norway's $1.5 trillion sovereign wealth fund, announced that it had lost about $92 million because a single date had been entered incorrectly in the calculation of its mandated benchmark [POB2401] (Financial Times, 2024).

Two highly sophisticated financial institutions, three decades apart, were affected by the same basic problem: a wrong value, entered by a presumably competent person, in a cell that nothing checked. The persistence of spreadsheet errors is the problem this paper takes up. Spreadsheet error has been studied in experimental work and field audits since the early 1990s; cell error rates of roughly one to five percent recur across the literature, inspection is unreliable, and developers are often overconfident about their own accuracy (Panko, 2003, 2008; Panko & Sprague, 1998; Powell, Baker & Lawson, 2008). Although audit tools, guidance and a substantial research community now exist, spreadsheet failures continue to occur and, as the corpus suggests, their consequences can still be serious.

Rather than begin with error theory, this paper begins with the stories themselves. The European Spreadsheet Risks Interest Group (EuSpRIG) has, for two decades, maintained a list of publicly reported spreadsheet disasters, the "horror stories". Individually, the cases are striking. Analysed together, cleaned and classified, they become evidence about how spreadsheets fail in practice. The paper asks four questions: why the same error mechanisms persist over decades (RQ1); what organisational conditions turn a routine error into a serious failure (RQ2); how the passage of time and the evolution of spreadsheets contribute to failure (RQ3); and what role invisibility and hidden structure play (RQ4). This paper contributes a cleaned and documented open corpus of 121 incidents, a closed twelve-category classification of the whole archive, and an account grounded in the worst cases of what these disasters consist of and why they have not stopped.

### 1.1 What the horror stories can and cannot tell us

The EuSpRIG archive is unusual among research materials: it is a running, public, decades-long record of how a single technology fails in the field, compiled as the failures were reported in the media. Quiet errors, near-misses caught in time, and disasters never traced to their spreadsheet do not appear, the corpus therefore contains mostly severe and publicly reported cases. Indeed, the stories contained in the corpus are likely the tip of the iceberg in the spreadsheet error landscape.

Experimental studies and field audits have shown that spreadsheets have cell error rates, the number of cells in error in comparison to the number of calculative cells in the spreadsheet, of roughly one to five percent and that the large majority of operational spreadsheets contain errors (Panko, 2008; Powell, Baker & Lawson, 2008). The horror stories show what happens when common spreadsheet errors are not discovered and meet real-world consequences.

### 1.2 How the corpus was built

The analysis set was assembled in two waves: the original EuSpRIG archive of 101 catalogued cases compiled from 1995 to early 2024, and a structured expansion of 28 cases reported between mid-2024 and 2026. Identifier prefixes (NS, FH, MPC, POB) cover the older cases and [ACST] cover the expanded cases. In the rest of the paper, cases are discussed and both a case reference is offered in square brackets, i.e. [ACST2611] and a citation to the reported error, often in the press in the normal way. A full table of all cases, sources and classification of those incidents can be found in Appendix A.

Of 129 catalogued cases, ten are set aside: Plimpton 322, a Babylonian arithmetic tablet retained only as a curiosity; a regulator's advisory that is a response to incidents rather than an incident; one entry that duplicates another's underlying event; two whose spreadsheet involvement is unconfirmed; one that aggregates dozens of breaches; and one deliberate contrast case reserved for the discussion.. The confirmed analysis set is N = 121. Cleaning resolved one identifier that had been assigned to two unrelated cases, normalised dates arriving in several formats, and recorded relationships between cases that describe the same event or cluster; all decisions are documented in the released dataset.

Each case was classified into exactly one of twelve failure categories (Table 1), assigned from the full case narrative. These categories are not all cell-level error types. Some, such as formula error or data entry, describe incorrect values. Others, such as hidden data, disclosure, governance and security exploitation, describe harm caused by a spreadsheet where no wrong calculation is present. Three cases whose sources name no mechanism are classified Unclassified rather than guessed.

The classification process combined AI-assisted thematic grouping with manual case-level review. First, the cases were gathered into a working table and passed to ChatGPT to produce an initial set of broad thematic groupings. These groupings were used as a starting point only. The authors then inspected the cases manually, case by case, checking the available details of each incident against the proposed groupings. Where a case had been misclassified, it was moved; where a grouping was too broad, ambiguous or overlapping, the category definitions were refined. This process produced the twelve failure categories used in the analysis. The refined table then became the basis for the rest of the analysis. The additional cases from mid-2024 to 2026 were subsequently assigned to these established categories according to the details available in the public reports.

## 2.0 RESULTS

The 121 incidents span 31 years and almost every domain in which spreadsheets are used: banking and markets, central and local government, health services, elections, defence, scientific research, sport and charity. Their consequences run from embarrassment through regulatory fines, billion-dollar restatements, corporate collapse, criminal proceedings and, in at least one case, death. Table 1 sets out the classification; Table 2 shows how the failure categories are distributed across reporting periods.

| Code | Failure category | Definition |
|---|---|---|
| FL | Formula & logic error | Wrong formula, reference, aggregation or model logic (sum vs count, omitted range, double counting, mistaken assumption). |
| DE | Data entry & transcription | Wrong value typed, transcribed, mislabelled or misplaced by a human. |
| DH | Data handling error | Error introduced moving or rearranging data: copy/paste, transfer between systems, partial sorts, merges, spreadsheet-to-system handoffs. |
| HD | Hidden or embedded data | Sensitive or material data structurally concealed within a shared file: hidden rows/columns/sheets, pivot caches, embedded templates, editable source layers. |
| DX | Disclosure & sharing error | The wrong file, or an unredacted one, is sent or published; nothing concealed and no value wrong; the disclosure itself is the failure. |
| CF | Capacity, format & conversion | The tool's own limits or silent conversions: row limits, truncation, auto-formatting, format coercion. |
| BL | Broken or incorrect links | Inter-sheet or inter-file link broken or mispointed. |
| DM | Deliberate manipulation & fraud | Spreadsheet features used intentionally to fabricate or conceal. |
| GV | Governance & overdependence | Failure of the surrounding process: uncontrolled sheets, no version or change control, key-person risk, systemic reliance. |
| DS | Design & usability failure | Structure misleads or defeats its users: complexity, hard coding, absent labels or template. |
| SX | Security exploitation | Spreadsheet used as an attack vector. |
| UN | Unclassified | No mechanism identifiable from available sources. |

Table 1. The twelve-category classification applied to the whole corpus. The complete case-by-case assignment is given in Appendix A.

| Failure category | 1995–2009 | 2010–14 | 2015–19 | 2020–26 | Total |
|---|---|---|---|---|---|
| Formula & logic (FL) | 2 | 14 | 11 | 8 | 35 |
| Data entry (DE) | 3 | 7 | 3 | 6 | 19 |
| Data handling (DH) | 4 | 3 | 4 | 6 | 17 |
| Hidden/embedded data (HD) | 1 | 1 | 2 | 8 | 12 |
| Governance (GV) | 3 | 3 | 1 | 3 | 10 |
| Capacity/format (CF) | 0 | 2 | 1 | 3 | 6 |
| Design/usability (DS) | 0 | 2 | 2 | 1 | 5 |
| Deliberate manipulation (DM) | 2 | 2 | 0 | 2 | 6 |
| Unclassified (UN) | 0 | 1 | 2 | 2 | 5 |
| Disclosure/sharing (DX) | 0 | 0 | 0 | 3 | 3 |
| Broken links (BL) | 1 | 1 | 0 | 0 | 2 |
| Security exploitation (SX) | 0 | 0 | 1 | 0 | 1 |
| Total | 16 | 36 | 27 | 42 | 121 |

Table 2. Failure category by reporting period (N = 121).

Two features stand out, and the rest of the paper is largely about them. The three categories describing ordinary spreadsheet work, namely formula and logic, data entry and data handling, together account for 71 of 121 incidents and appear in every reporting period without exception. One further category shows the clearest rise in reported cases across the whole period: hidden or embedded data. The sections that follow take the corpus apart along four analytical strands, each built around the cases that show the pattern most clearly.

### 2.1 Four analytical strands

The analysis answers these questions in four strands. It begins with the most common and persistent failures, formula, entry and handling errors (RQ1), before turning to hidden and embedded data as the clearest rising pattern in the corpus (RQ4). It then examines why some spreadsheet errors become serious organisational failures (RQ2), and finally considers how spreadsheets can carry dormant defects across time, reuse and handover (RQ3).

### 2.2 The mundane majority (RQ1)

The largest part of the corpus is also the most straightforward and recognisable. Seventy one of the 121 incidents are formula errors, data-entry slips and data-handling mistakes, the same broad error types described in spreadsheet literature (Panko, 2008). They recur across the whole corpus and at every level of institution. The point is clearest when cases are placed side by side. Fidelity's missing minus sign (1995) and Norway's mistyped date (2024), described above, are the same kind of execution slip at the same kind of institution, twenty-nine years apart. These ordinary errors resemble many other cases. In 2003 a TransAlta trader's cut-and-paste mistake misaligned the company's power-contract bids, which were binding and could not be withdrawn, costing $24 million the firm conceded was a simple clerical slip [NS10] (CIO, 2007). In 2026 Sutter County in California received about $105 million in state education money instead of the $25 million it was due, an extra $80 million, because a single row had shifted in the Excel file distributing the funds, an error the county caught and corrected only by noticing the windfall and handing it back [ACST2606] (Sacramento Bee, 2026).

When Virginia's alcohol authority mis-ran a lottery of more than 40,000 entries in 2023, its director of internal audit explained to the board that a sorting operation had been applied to only part of the data: *"It sorted some of it and didn't sort some of the rest"* [POB2305] (Moomaw, 2024), a simple mistake that could be present in any of the cases. Again in 2023, the Anaesthetic National Recruitment Office's (ANRO) scoring workbook, christened a "*Frankensheet*" for its absent template, missing labels and improvised structure, mis-aggregated interview scores so badly that every trainee anaesthetist in Wales was rendered *"unappointable"* [POB2309] (The Register, 2023).

In 2010 two Harvard economists, Carmen Reinhart and Kenneth Rogoff, published a finding that economic growth falls sharply once public debt passes 90% of GDP, a result widely cited by governments to justify austerity politics in the wake of the 2008 "credit crunch". Herndon *et al.* (2014) obtained the underlying spreadsheet and found that its averaging formula omitted five countries from the calculation [POB1302]. This omission sat alongside other methodological disputes, but its origin, a formula range that stopped short of the full data, is the same omission error found throughout this corpus, distinguished only by the influence of the conclusion it helped produce.

Most of these mistakes are human in origin, the slips and lapses of attention and the occasional mistake of reasoning that seem to occur whenever people work in spreadsheets.

## 2.3 The danger of hidden and embedded data (RQ4)

Of the twelve failure categories, hidden and embedded data shows the clearest rise in reported cases across the corpus. These are cases in which the visible spreadsheet may appear routine and legitimate, the damage comes from data and information unintentionally concealed in the spreadsheet. Hidden data is defined as data concealed from reasonable inspection, such as hidden rows, columns or worksheets. Embedded data refers to retained workbook material not visible in the presented output, such as pivot-table caches, source layers, embedded objects or templates.

In 2016 Blackpool Teaching Hospitals published spreadsheets whose pivot-table caches still held the personal data of thousands of staff; the file sat online for ten months and drew a £185,000 penalty [POB1603] (ICO, 2018). In 2023 the Police Service of Northern Ireland (PSNI) answered a routine freedom-of-information request by publishing a spreadsheet one of whose tabs contained a complete personnel download: the names, ranks and locations of all 10,799 officers and civilian staff, described in the coverage as a *"gold mine for terrorists"*, placing individuals at significant personal risk [POB2307] (ICO, 2024).

After the ICO issued explicit guidance in November 2023 warning public authorities to stop releasing original-source spreadsheets in response to Freedom of Information (FOI) requests, similar cases still occurred. South Gloucestershire Council published consultation responses relating to a planning application in a spreadsheet with hidden worksheets exposing 625 respondents' personal details [ACST2511] (BBC, 2025). Royal Cornwall Hospitals released an editable spreadsheet whose underlying layers held three years of staff sickness records [ACST2512] (Cornwall Live, 2025); Barking and Dagenham disclosed a file whose retained source data laid bare roughly 6,500 lines on people in homeless accommodation [ACST2603] (Yellow Advertiser, 2026). Cambridge University Hospitals, apologising for its own 2023 breach, described the mechanism clearly, the personal data shared had been "not immediately visible in the spreadsheet we provided" [POB2311] (Cambridge University Hospitals, 2023).

The most serious accidental leak of sensitive information in the corpus had already happened
breach, and the injunction itself, could not be reported for nearly two years, while it built and ran a covert resettlement programme for those judged to be in danger, at a cost reported to run from the hundreds of millions into the billions of pounds [ACST2507] (The Independent, 2025; National Audit Office, 2025).

The Ministry of Defence case shows the most severe accidental disclosure in the corpus. The spreadsheet contained information that was not visible to the sender but remained present in the workbook. The result was not a financial loss or a correction notice, but a covert resettlement programme for people whose exposure could place them at risk of death.

The same visibility problem also appears in deliberate cases of fraud, although the mechanism and intent differ. In these cases, spreadsheets are not accidentally carrying hidden information; they are used to create records that look legitimate while concealing falsehood. In 1999, quality-control workers at British Nuclear Fuels' Sellafield plant falsified the safety records for MOX nuclear-fuel pellets by cloning completed inspection spreadsheets and altering the lot and batch numbers, so that fuel of unknown mass and quality, some of it already shipped to Japanese reactors, appeared to have passed its secondary checks. Each doctored sheet looked legitimate on its own; the fraud surfaced only when an inspector noticed that successive lots held suspiciously similar data. The facility was suspended, the fuel repatriated, and some £40 million paid in compensation, and a public inquiry traced the episode less to the individuals than to the management decision to entrust safety-critical records to an unprotected spreadsheet in the first place [ACST1999BN] (Thorne, 2013).

The Allfirst trader John Rusnak concealed $691 million in losses for four years by manipulating spreadsheet data and persuading a supervisor that his paperwork need not be reconciled, and the fraud surfaced only when someone noticed the reconciliation had stopped [NS11] (Thorne, 2013). Even Bernie Madoff's $65 billion Ponzi scheme, which falls outside this corpus because the spreadsheet sustained the fraud rather than constituting it, relied on spreadsheets to manufacture a credible paper trail for regulators and investors. A Harvard professor found to have fabricated data behind a series of discredited research papers, ironically on the psychology of dishonesty, was exposed through auditing of the internal structure of the Excel files themselves [POB2306] (Simonsohn et al., 2023).

These cases should not be treated as a single type of failure. Some involve accidental disclosure; others involve deliberate falsification or concealment. What they share is that the visible spreadsheet did not reveal the full risk carried by the file. In accidental cases, the workbook retained information that should not have been shared. In deliberate cases, spreadsheets helped false records appear routine and trustworthy. In both, harm arose because the file was trusted without sufficient control over what it contained or represented.

## 2.4 Why some errors become horror stories (RQ2)

Most spreadsheet errors are likely to be corrected before they matter. Some, however, pass unnoticed by the modeller and by whatever checks or controls are in place, and are then used with the error intact. This becomes more likely when spreadsheets are built without prior planning, developed without a clear methodology, and released without systematic testing, a pattern documented by Powell et al. (2008) and Panko (2016). What distinguishes the cases that become horror stories is rarely the size of the error itself, but the criticality of the spreadsheet in which it occurs.

Some spreadsheets matter not because they calculate especially complex things, but because organisations use them to make binding decisions. They help decide elections, as in the 2024 UK general-election count in Putney, where a spreadsheet fault led officials to announce a total that was missing 6,558 votes [ACST2410] (BBC News, 2024). They set tax rates, as in Gallatin County, where a latent spreadsheet error caused taxpayers to overpay by $8.5 million and forced cuts to the fire department budget to reconcile the shortfall [ACST2509] (Mountain Journal, 2026). They distribute public money, as in Austin's 2026 payroll case, where a spreadsheet upload error caused the City of Austin to overpay 675 employees [ACST2607] (MSN, 2026). They determine careers, as in the ANRO workbook used to decide which doctors could be hired [POB2309] (The Register, 2023). They can also affect safety-critical infrastructure, as in the Edinburgh children's hospital case, where a design-spreadsheet error in ventilation specification meant the completed building was judged unsafe to open and required £16 million of remedial work [POB2002b] (BBC News, 2020a). In 2026, the Hobart Cup was run about 37 metres short of its advertised distance because the spreadsheet holding the starting-barrier positions omitted tacit knowledge about the real starting point for the 2400 metre course [ACST2602] (The Straight, 2026). These cases show that spreadsheet errors become serious when their outputs are acted on without sufficient testing.

Consequence depends on the context in which the spreadsheet is used, not simply on the type of error it contains. JPMorgan's 2012 "London Whale" model divided by a sum where it should have divided by an average, understating measured risk. The bank's own inquiry also recorded the surrounding conditions: a model approved under pressure, maintained through manual copy-and-paste, and inadequately resourced [POB1301] (JPMorgan Chase & Co., 2013). In 2018, Conviviality, a UK drinks group, made a £5.2 million arithmetic error in a spreadsheet; in an already fragile company, this contributed to profit warnings and collapse [POB1801] (Davies, 2018). By contrast, Snyder County in Pennsylvania made a similar kind of spreadsheet error in 2015, but the result was the discovery of roughly $330,000 of unexpected revenue, easing a budget shortfall [POB1506] (The Daily Item, 2015).

These cases show that the type of error does not determine the scale of harm. The outcome depends on the context, reach and the criticality of the spreadsheet: which decisions depend on it, who relies on its outputs, and what checks are applied before results are acted on. In that context, repeated official explanations of "human error" in the media, used at San Luis Obispo in 2017 [POB1705], Edinburgh in 2020 [POB2002b] and PSNI in 2023 [POB2307], are incomplete. They identify the immediate mistake, but not the organisational conditions that allowed it to be made, missed and eventually acted upon.

Appendix B summarises selected measurable consequences from the corpus, showing that these consequences include not only financial losses, fines and remediation costs, but also exposed populations, delayed public-health cases, distorted official declarations and epistemic harms.

### 2.5 Dormant spreadsheet bugs (RQ3)

Some spreadsheet errors behave less like one-off mistakes and more like dormant software bugs. They can be introduced early, remain inactive for years, and only become visible when the spreadsheet is reused, extended or placed under pressure. The Edinburgh hospital case is an example: the error was not unusual in mechanism, but it remained embedded in a working design spreadsheet for eight years before its consequences became visible. The South African pension fund case shows a similar pattern: a hard-coding error in an actuary's spreadsheet system reportedly caused substantial overpayments to exiting fund members over several years before the dispute reached the Pension Funds Adjudicator and Financial Services Tribunal [POB1905] (Independent Online, 2019).

The Hobart Cup case shows the same problem in a different setting. According to the published account, experienced staff knew that the barrier-position spreadsheet needed informal correction to ensure that the full distance was used in the race. When those staff left, the knowledge left with them. The spreadsheet remained, but the workaround did not, and the race was run short [ACST2602] (The Straight, 2026).

These cases show that spreadsheet risk is not only about how a spreadsheet is created. It is also about how it is maintained, reused and handed on. A spreadsheet can outlive the people who understand it. Like a software system, it can carry dormant defects that only surface when conditions change. People and tacit knowledge are also critical to this.

The same pattern appears across a whole research field. In 2016, an analysis of 35,175 supplementary Excel files found gene names corrupted throughout the published literature because Excel automatically converted symbols such as SEPT1 into dates [POB1605] (Ziemann, Eren & El-Osta, 2016). In 2020, the body responsible for naming human genes responded by renaming some genes to avoid the Excel conversions [POB2003] (Bruford et al., 2020). This was a remarkable outcome. It proved easier to change the naming of human genes than to remove a spreadsheet default from scientific practice. A 2023 follow-up found that the errors were still appearing [POB2308] (Abeysooriya et al., 2021).

The corpus also includes one useful contrast case. In 2025, the Williams Formula One team retired the 20,000-cell Excel workbook that had been running its car-build process. The team replaced it with engineered systems after the workbook's limits had contributed to operational problems. Williams did something that many organisations in this corpus did not do. It recognised that a spreadsheet had become critical infrastructure before it produced a public disaster, and it treated it accordingly.

### 2.6. A note on classifying these failures

It is natural to ask how this corpus relates to the established taxonomies of spreadsheet error, the best known of which divides quantitative errors into mechanical slips, logic errors and omissions, later extended to separate innocent errors from deliberate violations (Panko & Halverson, 1996; Panko & Aurigemma, 2010). The relationship is illuminating chiefly where it breaks down. For the mundane majority the older scheme fits comfortably: data-entry cases are mechanical slips, the JPMorgan, Reinhart–Rogoff and West Coast Main Line rail-franchise cases are logic or model-logic errors, the omitted vote tallies are omissions, and the fraud cases (Allfirst, the Clallam cashier, the research fabrications) are violations rather than errors at all. The West Coast Main Line rail-franchise case illustrates the boundary between spreadsheet error and spreadsheet-enabled governance failure. Reported as involving spreadsheet modelling errors in the Department for Transport's procurement process, the mechanism concerned a flawed forecast/risk model in which inflation, passenger-number assumptions and bidder risk guarantees were evaluated incorrectly. The failed £9bn procurement process led to refunds of around £40m to the four bidders, with total taxpayer costs reported as potentially rising to £300m [ACST1201] (Brooks, 2012). The case is coded as FL, Formula & logic error because the available account points to erroneous model logic and assumptions, while the wider incident also shows how spreadsheet models can acquire public-procurement authority.

But a large share of the corpus falls outside any taxonomy built to classify wrong numbers, because the cases do not depend on an incorrect computed value. Counting the hidden-data exposures, the disclosure errors, the capacity and format failures, the governance breakdowns and the security exploit, 32 of the 121 incidents, more than a quarter, fall into categories where no incorrect calculation need be present. In the PSNI case, for example, no incorrect calculation was identified; the harm lay in what the workbook disclosed.

The implication is that public spreadsheet incidents cannot be understood only as failures of calculation. In many cases the spreadsheet is harmful because it discloses information, carries hidden structure, supports weak governance or creates a security exposure. Table 3 summarises how the corpus maps onto the established error-taxonomy tradition.

| Corpus categories | Established-taxonomy correspondence | Note |
|---|---|---|
| DE, DH (36) | Mechanical slips | Operation-level handling errors fit only loosely in a cell-by-cell scheme. |
| FL (35) | Logic, mechanical (pointing) and omission | Splits three ways; case-level assignment is contestable and is given in the dataset, not asserted here. |
| DM (5) | Violations, not errors | Intentional fabrication or concealment. |
| DS (5) | Qualitative errors | Defects that breed future wrong values rather than being wrong themselves. |
| HD, DX, CF, GV, SX (32) | Outside the taxonomy | More than a quarter of the corpus: no incorrect computed value need be present. Risk has expanded beyond calculation error. |

Table 3. The corpus mapped onto the established error taxonomies. The fit is good for the mundane majority and absent for over a quarter of cases.

### 3.0 Discussion: What organisations can do

The mundane majority persists because data-entry, formula and handling errors are easy to make and hard to find (Panko, 2008; Panko & Sprague, 1998). Spreadsheet research shows that even careful users still make mistakes, and checking one's own work is not enough to remove them. There is no realistic level of individual care that will drive spreadsheet error rates to zero, and thirty years of spreadsheet horror stories underline this. Telling people to "be more careful" or "do not make mistakes" has effectively been the policy of many organisations for decades, and it has not worked. The recommendations below follow from the failure categories identified in this corpus.

### 3.1 Build in basic checks before relying on spreadsheet outputs

The cases in Section 2.2 show that many spreadsheet disasters were not failures of advanced modelling, but failures of elementary control. Before a spreadsheet result is used, someone should ask whether the row count is plausible, whether all expected records are present, whether signs, dates and units make sense, whether subtotals reconcile to the final total, whether the whole budget or allocation has been accounted for, and whether the output agrees with an independent source. These checks are evidently missing from many cases in this corpus. They are also the kinds of checks that would catch common mistakes such as a missing minus sign, a mistyped date, a row moving out of place, an omitted range, or a copy-and-paste error before the spreadsheet is used to make a decision. The case for systematic spreadsheet testing, self-checking models, cross-foot checks, and independent reconciliation has been made repeatedly within this field (Pryor, 2004; Nash & Goldberg, 2005; O'Beirne, 2005; Ayalew, *et al.*, 2008; Panko, 2006).

The Crypto.com transfer of A$10.5 million went unnoticed for seven months not because the error was subtle, but because nothing compared the sum paid with the refund intended. The same pattern appears in the Westminster City Council budgets that became deficits and funding files that delivered the wrong amount. A reconciliation step does not require the developer to be more careful or more skilled. It assumes that an error may already have happened, and asks whether the numbers leaving the spreadsheet agree with something computed, recorded, or expected independently of it.

### 3.2 Treat fraud as a different threat model

The fraud cases in the corpus require a different response from the accidental-error cases. Reconciliation and testing help when a spreadsheet is wrong by mistake, but fraud involves someone deliberately making the spreadsheet look right. In these cases, the problem is not only that a value may be wrong, but that the workbook may have been designed to conceal the fact. Panko and Ordway (2005) make this distinction in the Sarbanes–Oxley context, and Mittermeir et al. (2008) argue that fraudulent spreadsheet faults must be found by auditors without relying on the cooperation of the person who created them. The practical controls are therefore different: restricted access, protected formulas, audit trails, version history, segregation of duties, independent review, and reconciliation against external records. The Allfirst case shows why these matter. The spreadsheet did not merely contain false numbers; it helped sustain a false account of trading activity until reconciliation failures were finally noticed (Butler, 2009).

### 3.3 Match scrutiny to criticality, not to size

Section 2.4 showed that the consequence of a spreadsheet error is set not by the complexity of the spreadsheet, but by its criticality, the importance of the decision that it makes or supports. Several of the most serious failures involved relatively ordinary workbooks whose outputs were used to set tax rates, allocate votes, distribute payroll or specify hospital ventilation. A small spreadsheet that makes a binding decision deserves more assurance than a large model whose output informs little.

Appendix B groups selected cases by reported consequence type, showing both the range and magnitude of harms that follow when spreadsheet errors reach public or organisational decisions. The table shows that financial harm itself takes several forms: direct losses, erroneous payments, accounting and valuation distortions, regulatory fines, compensation, remediation costs and disputed or potential exposures. Some individual cases are measured in tens or hundreds of millions, and the broader corpus includes reported consequences that reach into the billions. However, many consequential cases are not primarily financial. Spreadsheet incidents can expose thousands of people or records, compromise safety-critical information, delay public-health action, distort official declarations, affect appointments or employment processes, and damage scientific or public knowledge. The financial cases are often easiest to quantify, but they are not necessarily the most serious.The Ministry of Defence Afghan breach, the PSNI disclosure, the Public Health England Covid case and the gene-name conversion errors show that spreadsheet failures can also affect personal safety, institutional trust, public health, scientific validity and democratic or administrative authority.

The problem is not simply that spreadsheets contain errors. It is that spreadsheets are used for almost everything, including tasks for which a controlled database, workflow system or engineered application would often be more appropriate. Panko and Port (2013) describe end-user computing as the “dark matter” of corporate IT: enormous in quantity and importance, but largely invisible to formal IT governance, management and information-systems research. The cases examined here show the consequence of that invisibility. Spreadsheets appear as election calculators, payroll interfaces, hospital-design records, public-disclosure mechanisms and fraud-enabling artefacts, often without being treated as operational systems until after they fail.

The cases therefore suggest that review should be based on consequence rather than apparent complexity. A spreadsheet may look routine, but if it sets a tax rate, allocates votes, distributes payroll or specifies hospital ventilation, it deserves close scrutiny. This risk-centred approach is consistent with the spreadsheet-control literature, which argues that organisations should identify critical spreadsheets, assign ownership, apply version and change control, and target review effort at the spreadsheets whose failure would matter most (Chambers & Hamill, 2008; Glass et al., 2009; Rittweger & Langan, 2010; Lemon & Ferguson, 2010; Turner, 2019).

### 3.4 Engineer the act of sharing

The hidden-data and disclosure cases in Section 2.3 are not caused by faulty calculations. They occur because a workbook is shared in its original format and contains information that the sender cannot see or does not realise is still present. The PSNI, South Gloucestershire, Royal Cornwall and Ministry of Defence breaches could each have been avoided by exporting the relevant data to a flat format such as CSV, or to a generated PDF, rather than sharing the original workbook. This would make the released file match the visible output and remove hidden rows, worksheets, pivot caches and other retained material. The ICO has already recommended this approach for safe disclosure (Information Commissioner’s Office, 2023). The fact that breaches continued after the warning shows that guidance alone is insufficient. Organisations need release processes that prevent or restrict the external sharing of raw workbooks, rather than relying on users to remember not to send them.

### 3.5 The coming test: AI-assisted spreadsheets

Spreadsheets are increasingly built and modified with generative AI, but there is little evidence that machine assistance will break the failure patterns documented in this corpus. Existing work suggests instead that large language models reproduce familiar spreadsheet risks. Thorne (2023) found that ChatGPT can generate correct formulae in favourable conditions, but breaks down when requirements are limited, uncertain or complex, producing confident but wrong outputs. O'Beirne (2023) therefore recommends verification before trust. Later benchmarking reached a similar conclusion: leading models handle simple spreadsheet tasks well but falter on complex, multi-step operations, often producing plausible but incorrect results (Thorne, 2025), a pattern echoed by Dong et al. (2024). Grossman et al. (2025) similarly show that GPT-based spreadsheet modelling tools must be assessed not only for whether they produce an answer, but for whether they produce reusable spreadsheet models with inputs, formulas, labels and accuracy intact. AI does not eliminate the mundane errors described in Section 2.2, it can generate them.

The greater risk is that AI changes the conditions of authorship. It lets users produce more spreadsheet content faster, makes hidden structure harder to inspect when formulas are generated rather than understood, and may weaken the human judgement on which informal spreadsheet control often depends. It may also change the pattern of error. AI may reduce some small slip errors, such as mistyped syntax or simple transcription mistakes, but it can also create systematic errors that are repeated consistently. If an AI-generated formula is right, it may be right everywhere it is reused. If it is wrong, it may be wrong everywhere as well. Sarkar et al. (2024) argue that the principal risk is not hallucination alone but the erosion of critical thinking as users delegate judgement to trusted systems; Lee et al. (2025) similarly find that confidence in AI is negatively associated with critical thinking in AI-assisted tasks. The Hobart Cup case shows the fragility of tacit knowledge when it is not formalised (Section 2.5). AI-assisted authorship risks removing that understanding from the outset.

AI therefore extends rather than retires the corpus. AI-assisted spreadsheets inherit the same risks: wrong formulas, hidden structures, corrupted handoffs and misplaced authority. They may also provide new defences, including more reliable calculation components and "co-audit" interfaces that help users find and correct generated errors (Sarkar et al., 2024). Whether AI becomes a new source of spreadsheet horror stories or a meaningful intervention will depend less on model accuracy alone than on whether tools preserve the understanding and critical judgement of the person nominally in control.

### 3.6 A programme of mitigation in depth

No single measure addresses the risks presented in the whole corpus, because the corpus is not a single kind of failure. The mitigations above map onto the strands identified in the analysis. Reconciliation catches many ordinary errors. Fraud controls address deliberate manipulation. Scrutiny matched to criticality limits the damage when a small sheet carries a large consequence. Engineered sharing closes the hidden-data and disclosure routes. These measures are complementary rather than alternatives. An organisation serious about not becoming the next entry in this archive would adopt them as layers.

None of these measures is novel, expensive or technically demanding. Each has been articulated within this community over two decades (Croll, 2009; Panko & Ordway, 2005). The problem is not lack of knowledge, but the persistent treatment of spreadsheets as trivial office documents rather than as the operational infrastructure the corpus shows them to be. The first and most consequential step an organisation can take is simply to identify which of its spreadsheets are critical and to stop treating them as disposable. The recommendations above then follow naturally, and the rest is a matter of building the layers in before, rather than after, the artefact fails. In practice, many organisations still appear not to identify spreadsheets as potential sources of critical operational risk.

### 3.7 Limitations

This corpus is a public record, not a representative sample of all spreadsheet errors. It captures cases that were reported, traceable to spreadsheets and serious or visible enough to enter the public record. It therefore over-represents public, severe and institutionally consequential incidents, and under-represents quiet errors, near misses, internal corrections and cases where the spreadsheet mechanism was never identified. Many source reports provide only limited detail, so the classification is necessarily case-level rather than cell-level. Where the available sources did not identify a mechanism, the case was left unclassified rather than inferred.

### 3.8 Future work

The cleaned corpus creates a basis for future work as well as a snapshot of current knowledge. The immediate next step is to make the EuSpRIG horror stories archive more searchable and usable: allowing users to filter cases by date, sector, failure category and consequence, and to move from the summary classification to the underlying story and source. A second line of work is methodological: comparing this public corpus against audit datasets and organisational case studies, where available, to test whether the patterns reported here also appear in less public spreadsheet failures. A third line concerns AI-assisted spreadsheets, where future horror stories may arise not from typing mistakes alone but from generated formulas, generated models and generated confidence.

### 3.9 Conclusion

The cases in this corpus show that the basic causes of spreadsheet failure have changed little over time. The same kinds of mistakes appear across the whole period: a missing sign, a mistyped date, a mis-sorted range, a hidden tab or retained data inside a shared file. What varies is the setting in which those mistakes occur. A routine spreadsheet error may cause little harm in one context, but serious consequences in another if the workbook is used to allocate money, report financial results, support public services, decide appointments, disclose data or manage safety-critical work. In the most serious case in this corpus, hidden rows in a Ministry of Defence spreadsheet contributed to a superinjunction and a covert evacuation programme. The problem is therefore not simply that spreadsheets contain errors, but that organisations often use them for important decisions without treating them as important systems. The corpus shows that small errors can still reach high-stakes decisions, and that hidden-data and disclosure incidents are becoming more visible and more serious. The consequence groups in Appendix B underline that these high-stakes decisions produce multiple forms of harm, only some of which are financial. Organisations therefore need to treat critical spreadsheets as operational infrastructure, with ownership, testing, version control, release controls and review proportionate to the decisions they support. The cleaned corpus and classification offered in this paper are intended to help make those risks easier to see, compare and prevent.

## Appendix A. The classified corpus

All cases in chronological order, with the failure-category code from Table 1. Expansion-wave cases (2024–2026) carry the prefix ACST; earlier identifiers preserve the EuSpRIG archive's original prefixes. The Source column links to a primary report for each case where one is available; the complete set of sources and all coding fields are in the released dataset.

| ID | Year | Code | Title | Source |
|---|---|---|---|---|
| NS13 | 1995 | DE | Fidelity Magellan dividend misstatement: omitted minus sign ($1.3bn) | source |
| NS12 | 1998 | GV | National financial-reporting failure linked to poor spreadsheet controls | source |
| ACST1999BN | 1999 | DM | BNFL Sellafield: MOX nuclear-fuel safety data falsified by cloning spreadsheets and altering lo | source |
| NS11 | 2002 | DM | Allfirst/AIB trading fraud via manipulated spreadsheet links ($700m) | source |
| NS09 | 2003 | FL | Fannie Mae spreadsheet computation error under new accounting standard ($1.2bn) | source |
| NS10 | 2003 | DH | TransAlta cut-and-paste error in binding power-contract bids ($24m) | source |
| NS08 | 2004 | FL | University of Toledo budget: formula projected enrolment rise not fall ($2.4m forecast gap) | source |
| NS07 | 2005 | DE | RedEnvelope cost forecast wrong from a single incorrect cell entry | source |
| NS06 | 2005 | DH | Virginia Tech regional study understated figures via cut-and-paste error (11% vs 20%) | source |
| NS04 | 2005 | HD | Westpac: unreleased results exposed via improperly hidden template data | source |
| NS05 | 2005 | DE | Kodak severance accrual error: extra zeros added ($11m overstatement) | source |
| NS03 | 2006 | GV | Nevada City budget: spreadsheet error misstated deficit, later corrected | source |
| NS02 | 2008 | GV | Credit Suisse: over-reliance on complex trading spreadsheets (£5.6m fine) | source |
| NS01 | 2009 | DH | Revenue misreported after faulty data transfer into spreadsheet (~11%) | source |
| FH1231 | 2009 | DH | Foutje in spreadsheet kost waterschap 2,7 miljoen | source |
| FH1205 | 2009 | BL | $15 million mistake: that representative doesn't work for the company anymore | source |
| FH1209 | 2010 | FL | Spreadsheet error costs time and money, yet again | source |
| FH1206 | 2010 | UN | $153 million error in Department of Energy Tax spreadsheet | source |
| FH1208 | 2010 | DE | Flintshire County Council school cash blunder 'down to spreadsheet error' | source |
| FH1229 | 2010 | DH | Blatant Data Error At The Federal Reserve | source |
| FH1223 | 2010 | CF | Millions undisclosed by TX congressman | source |
| FH1234 | 2010 | FL | Ex-county workers consider suit over reduced retirement | source |
| FH1235 | 2010 | FL | NAMB corrects missionary count data to GCR Task Force | source |
| FH1222 | 2010 | DS | Hungary's Central Services Directorate fined for misleading spreadsheet | source |
| FH1211 | 2010 | FL | Council 'loses' £21m spreadsheet error | source |
| FH1224 | 2010 | DE | Light Dept. Memo Defends Substation Team, Cost Increases | source |
| FH1212 | 2010 | DE | Provision for income taxes understated by $36.4 million | source |
| FH1210 | 2011 | DS | $1M went missing as staff managed "monstrous spreadsheets." | source |
| FH1213 | 2011 | DH | County overlooks, then finds, taxable property worth $1.26 billion | source |
| FH1217 | 2011 | DH | 1,791 voters inflated to 4,870 | source |
| FH1220 | 2011 | GV | $350,000 fine for a spreadsheet error | source |
| FH1201 | 2011 | GV | Securities and Exchange committee itself relies on spreadsheets. | source |
| FH1215 | 2011 | CF | MI5 makes 1,061 bugging errors | source |
| FH1207 | 2011 | DM | Clallam County cashier hides rows in a spreadsheet to cover up theft | source |
| FH1221 | 2011 | FL | State Auditor of Kentucky: bad smells from Sanitation District #1 | source |
| FH1228 | 2011 | DE | Major Blunder By Norwegian Central Bank Gives People Totally Wrong Impression About Interest Ra | source |
| FH1216 | 2011 | FL | £4.3M spreadsheet error leads to resignation of its chief executive | source |
| FH1202 | 2011 | BL | Bad spreadsheet link: $6 million error and $12,500 audit fee | source |
| FH1219 | 2011 | DE | Test errors impede History applications | source |
| FH1218 | 2011 | FL | Interest calculation low by $400,000 | source |
| FH1226 | 2011 | FL | Computer errs in tally | source |
| FH1227 | 2011 | FL | King's Fund apologise for "error" in figures on health spending in Wales | source |
| FH1225 | 2011 | GV | AXA Rosenberg paying $242M to settle case on code | source |
| FH1204 | 2012 | DE | Spreadsheets behind Olympic data misentry | source |
| MPC02 | 2012 | DM | UBS trader discussed "slush account" with Adoboli | source |
| FH1203 | 2012 | HD | AstraZeneca spreadsheet slip-up | source |
| MPC01 | 2012 | FL | Calculation error leaves school short by $25 million | source |

| ACST1201 | 2012 | FL | West Coast Main Line rail franchise: flawed spreadsheet forecast/risk model in public procurement; £40m bidder refunds and possible £300m taxpayer cost | source |
|---|---|---|---|---|
| POB1302 | 2013 | FL | Researchers embarrassed after student finds error | source |
| POB1301 | 2013 | FL | Report identifies lack of spreadsheet controls | source |
| MPC03 | 2013 | DE | Brownback budget director apologizes for incorrect figures | source |
| MPC04 | 2014 | FL | Spreadsheet Error Costs Tibco Shareholders $100M | source |
| POB1506 | 2015 | FL | Spreadsheet error helps Snyder County budget | source |
| POB1507 | 2015 | FL | Spreadsheet error overstated oil production from Fort Berthold Reservation | source |
| POB1504 | 2015 | DS | Censured after pointing out errors | source |
| POB1503 | 2015 | DH | SFMTA Retracts Report of 651% Jump in Bike Traffic | source |
| POB1501 | 2015 | FL | Officials: Whistle-blower case spreadsheet error | source |
| POB1502 | 2015 | FL | Local Government Association refund error | source |
| POB1505 | 2015 | FL | Tax flub raises budget questions | source |
| POB1601 | 2016 | SX | BlackEnergy .XLS Dropper | source |
| POB1606 | 2016 | FL | SolarCity adviser Lazard made mistake in Tesla deal analysis | source |
| POB1603 | 2016 | HD | Blackpool Teaching Hospitals fined by Data Protection Commissioner | source |
| POB1605 | 2016 | CF | Gene name errors are widespread in the scientific literature | source |
| POB1604 | 2016 | FL | M&S results hit by spreadsheet mishap | source |
| POB1607 | 2016 | FL | Wisconsin presidential recount will cost $3.9 million | source |
| POB1705 | 2017 | DE | SLO County budget surplus now a deficit after 'human error' | source |
| POB1704 | 2017 | DH | Risks of using a spreadsheet for election calculation | source |
| POB1706 | 2017 | UN | Eskom made R1.5bn spreadsheet error in Optimum fine | source |
| POB1702 | 2017 | DH | Spreadsheet errors cost Clallam $494,157 | source |
| POB1703 | 2017 | HD | Emailed spreadsheet contained private data in 'hidden' columns | source |
| POB1701 | 2017 | GV | Unofficial spreadsheets land Italian pharma plant with regulatory warning | source |
| POB1802 | 2018 | UN | Buskers Fest lost money last year due to 'spreadsheet error' | source |
| POB1801 | 2018 | FL | Drinks company £5.2M spreadsheet arithmetic error | source |

| POB1905 | 2019 | DS | “The error constituted maladministration on the part of the actuary” | source |
|---|---|---|---|---|
| POB1904 | 2019 | FL | AG: State overpaid Stroudsburg nearly $500K | source |
| POB1906 | 2019 | DE | State fund for jobs loses €750k due to ‘human error’ | source |
| POB1902 | 2019 | DE | A company submitted a bid nearly $3 million lower than it should have been | source |
| POB1903 | 2019 | DH | NYCLU apologizes for misstating racial disparity in Schenectady pot arrests | source |
| POB1901 | 2019 | FL | A week after posting earnings, Canopy Growth restates EBITDA after calculation error | source |
| POB2002a | 2020 | DH | ‘Spreadsheet formula’ to blame for over $16,000 in property tax errors | source |
| POB2001 | 2020 | CF | Data not controlled, 16000 UK Covid-19 test results lost for a week | source |
| POB2003 | 2020 | CF | Scientists rename human genes to stop Microsoft Excel from misreading them as dates | source |
| POB2002b | 2020 | DE | Spreadsheet error led to Edinburgh hospital opening delay | source |
| POB2106 | 2021 | DE | PRA fines Standard Chartered Bank £46,550,000 | source |
| POB2101 | 2021 | GV | PRA fines Metro Bank £5,376,000 for failing in its regulatory reporting governance and controls | source |
| POB2104 | 2021 | HD | Nothing is really hidden in Excel | source |
| POB2103 | 2021 | GV | The so-called “Highly Regulated Spreadsheet” | source |
| POB2102 | 2021 | GV | Canada Emergency Wage Support (CUSS) spreadsheet ‘riddled with errors’ | source |
| ACST2507 | 2022 | HD | UK MoD Afghan resettlement hidden-rows spreadsheet breach | source |
| ACST2201CR | 2022 | DE | Crypto.com refund: account number entered in payment-amount field; A$10.5m sent instead of A$10 | source |
| POB2202 | 2022 | DH | Police in Finland lose hacked therapy centre criminal reports after spreadsheet error | source |
| POB2301 | 2022 | DM | Generated fake data detected by coincidence with Excel limit | source |
| POB2302 | 2023 | FL | Formula error inflated myocarditis statistics | source |
| POB2307 | 2023 | HD | PSNI data breach of “monumental proportions”, personal data of 10,799 staff | source |
| POB2306 | 2023 | DM | Fraud in four Harvard Business school papers by Excel manipulation | source |
| POB2309 | 2023 | DS | ‘Frankensheet’ sorting errors hamper recruitment | source |
| POB2303 | 2023 | FL | Matt Parker broadcast on Spreadsheet disasters | source |
| POB2304 | 2023 | HD | ‘Spill’ of ‘hidden’ Excel rows derails Proud Boys trial | source |
| POB2305 | 2023 | DH | Blame Excel, not the user | source |
| POB2308 | 2023 | CF | Gene name errors still there, 20 years on | source |
| POB2311 | 2023 | HD | Cambridge University Hospitals NHS apologises for data breach | source |
| POB2401 | 2024 | DE | The Norwegian Sovereign Wealth Fund's $92 Million Excel Error | source |
| ACST2411 | 2024 | DX | Northern Ireland Department of Education SEN event spreadsheet data breach | source |
| ACST2410 | 2024 | FL | Putney general election spreadsheet issue omits 6,558 votes from announcement | source |
| ACST2504 | 2024 | FL | University of Arizona $240m cash-on-hand forecasting mistake | source |
| ACST2412 | 2024 | FL | Richmond Park general election Green Party result miscalculated after spreadsheet formula error | source |
| ACST2501 | 2025 | DX | Colorado veterans' personal information leaked in email attachment mistake | source |
| ACST2508 | 2025 | DH | Tarrant Appraisal District vote allocation spreadsheet error | source |
| ACST2511 | 2025 | HD | Local authority publishes hidden worksheets with consultation respondents' details | source |
| ACST2512 | 2025 | HD | Royal Cornwall Hospitals NHS Trust editable spreadsheet FOI breach | source |
| ACST2509 | 2025 | FL | Gallatin County Excel formula causes Big Sky taxpayers to overpay | source |
| ACST2503 | 2025 | UN | Pine Crest Nursing Home financial spreadsheet correction | source |
| ACST2502 | 2025 | FL | Canton of Thurgau building programme Excel error costs millions | source |
| ACST2611 | 2026 | UN | North Lincolnshire roads ratings confusion from spreadsheet interpretation error | source |
| ACST2612 | 2026 | DE | Significant payroll errors from manual spreadsheet adjustments | source |
| ACST2609* | 2026 | FL | Porirua/Wellington Water wastewater tank budget increase linked to spreadsheet miscalculation | source |
| ACST2610* | 2026 | GV | NERSA/Eskom R76bn power-bill error dispute | source |
| ACST2608* | 2026 | BL | Hawai'i LNG business case built on broken spreadsheet | source |
| ACST2605 | 2026 | FL | Aquarion deal challenged after spreadsheet/math error | source |
| ACST2606 | 2026 | DH | Sutter County returns $80m extra education funds after Excel row error | source |
| ACST2607 | 2026 | DH | Austin overpays 675 employees after payroll spreadsheet upload error | source |

| ACST2602 | 2026 | DE | Hobart Cup run short after barrier spreadsheet error | source |
|---|---|---|---|---|
| ACST2601 | 2026 | DX | Hull school sends new intake spreadsheet to parents | source |
| ACST2603 | 2026 | HD | Barking and Dagenham homelessness accommodation spreadsheet leak | source |

*Table A1. The complete corpus and its classification. 'source' links open the primary report for that case; a dash indicates no single linkable source (see dataset).*

**Appendix B Consequences of Spreadsheet errors**

Table B1 groups selected cases by reported consequence type. It is not an exhaustive coding of every case in the corpus. Rather, it summarises cases where public reporting identifies a clear financial amount, affected population, operational count, or other measurable consequence.

| Consequence group | Examples from corpus | Reported scale |
|---|---|---|
| Financial loss and overpayment | TransAlta [NS10]; Norwegian sovereign fund [POB2401]; Crypto.com [ACST2201CR]; Sutter County [ACST2606]; Austin payroll [ACST2607]; Gallatin [ACST2509] | Hundreds of thousands to hundreds of millions |
| Financial reporting, valuation and market distortion | Fidelity [NS13]; Fannie Mae [NS09]; TIBCO [MPC04]; Conviviality [POB1801]; M&S [POB1604] | Millions to billions |
| Fines, compensation and remediation | Standard Chartered [POB2106]; Metro Bank [POB2101]; Credit Suisse [NS02]; West Coast Main Line [ACST1201]; Edinburgh hospital [POB2002b]; BNFL [ACST1999BN] | Millions to potentially billions |
| Potential or disputed exposure | West Coast Main Line [ACST1201]; Aquarion [ACST2605]; NERSA/Eskom [ACST2610]; Hawaiʻi LNG [ACST2608] | Hundreds of millions or more |
| Privacy and security exposure | MoD Afghan [ACST2507]; PSNI [POB2307]; Colorado VA [ACST2501]; South Gloucestershire [ACST2511]; Barking and Dagenham [ACST2603]; Cambridge NHS [POB2311] | Thousands to tens of thousands of people or records |
| Public-health and safety consequence | PHE Covid [POB2001]; Edinburgh hospital [POB2002b]; BNFL [ACST1999BN]; Hobart Cup [ACST2602] | 16,000 Covid cases delayed/omitted; safety-critical records or infrastructure affected |
| Democratic and administrative authority | Putney [ACST2410]; Richmond Park [ACST2412]; ANRO [POB2309]; Tarrant [ACST2508]; Gallatin [ACST2509] | Thousands of votes; hundreds of employees, candidates or taxpayers affected |
| Epistemic and scientific harm | Gene-name errors [POB1605; POB2308]; Reinhart–Rogoff [POB1302]; myocarditis statistics [POB2302]; Harvard/Gino [POB2306] | Often not meaningfully monetisable |